\documentclass[a4paper]{article}
\usepackage[]{natbib}
\usepackage[fleqn,intlimits]{amsmath}
\usepackage{amsfonts,amssymb,bm}
\usepackage{graphicx,float}
\usepackage[left=2cm,right=2cm,top=2cm,bottom=2cm]{geometry}
\usepackage[margin=1cm,font=small,textfont=it,labelfont=bf,labelsep=endash]{caption}
\usepackage{color}

\begin{document}
\title{Nonstationary filtered shot-noise processes and applications to neuronal membranes}
\date{September 13, 2015}
\author{Marco Brigham, Alain Destexhe}
\maketitle
\begin{abstract}

Filtered shot noise processes have proven to be very effective in modeling the evolution of systems exposed to
shot noise sources and have been applied to a wide variety of fields ranging from electronics through biology. In
particular, they can model the membrane potential $V_m$ of neurons driven by stochastic input, where these filtered
processes are able to capture the nonstationary characteristics of $V_m$ fluctuations in response to presynaptic input
with variable rate. In this paper we apply the general framework of Poisson point processes transformations to
analyze these systems in the general case of nonstationary input rates. We obtain exact analytic expressions,
as well as different approximations, for the joint cumulants of filtered shot noise processes with multiplicative
noise. These general results are then applied to a model of neuronal membranes subject to conductance shot
noise with a continuously variable rate of presynaptic spikes. We propose very effective approximations for the
time evolution of the $V_m$ distribution and a simple method to estimate the presynaptic rate from a small number
of $V_m$ traces. This work opens the perspective of obtaining analytic access to important statistical properties of
conductance-based neuronal models such as the first passage time.

\end{abstract}

\section{Introduction}
We investigate the statistical properties of systems that can be described by the filtering of shot noise input through a linear first-order ordinary differential equation (ODE) with variable coefficients. Such systems give rise to filtered shot noise processes with multiplicative noise. The membrane potential $V_m$ fluctuations of neurons can be modeled as filtered shot noise currents or conductances \citep{verveen1974membrane, tuckwell1988introduction}. These fluctuations have been previously analyzed in the stationary limit of shot noise conductances with constant rate \citep{kuhn2004neuronal,richardson2005synaptic,rudolph2005extended,burkitt2006review1}, and an exact analytical solution has been obtained for the mean and joint moments of exponential shot noise \citep{wolff2008method, wolff2010mean}. However, many neuronal systems evolve in nonstationary regimes driven by shot noise with variable input rate. A typical example is provided by visual system neurons that receive presynaptic input with time-varying rate that reflects an evolving visual landscape. Modeling studies often consider the exponential shot noise case, whereas biological systems may display larger diversity including slow rising impulse response functions similar to alpha and bi-exponential functions, for example. Previous studies have addressed nonstationary exponential shot noise conductances and nonstationary currents \citep{cai2006kinetic, amemori2001gaussian, burkitt2006review2}.\\
\\
Poisson point processes (PPP) provide a natural model of random input arrival times that are distributed according to a Poisson law that may vary in time. Application-oriented treatments of PPP theory and PPP transformations can be found in Refs.~\citep{moller2003statistical, streit2010poisson}. The key idea of this article is to express the filtered process as a transformation of random input arrival times and to apply the properties of PPP transformations to derive its nonstationary statistics. Using this formalism we derive exact analytical expressions for the mean and joint cumulants of the filtered process in the general case of variable input rate. We develop an approximation based on a power expansion of the expectation about the deterministic solution. We apply these results to a simple neuronal membrane model of sub-threshold membrane potential $V_m$ fluctuations that evolves under shot noise conductance with continuously variable rate of presynaptic spikes.\\
\\
Shot noise processes are simple yet powerful models of stochastic input that correspond to the superposition of impulse responses arriving at random times according to a Poisson law. Systems evolving under shot noise input have been observed across many domains, such as electronics \citep{campbell1909study, schottky1918spontane}, optics \citep{picinbono1970photoelectron, rousseau1971statistical}, and many other fields \citep{snyder1991random, parzen1999stochastic}. Shot noise was discovered in the early works of Campbell and Schottky \citep{campbell1909study, schottky1918spontane}. Key theoretical results were obtained by Rice \citep{rice1945mathematical} and a modern review of their probabilistic structure is presented in Ref.~\citep{rice1977generalized}. Filtered shot noise processes with multiplicative noise are an extension of \emph{filtered Poisson process} \citep{snyder1991random, parzen1999stochastic, streit2010poisson} that are generated by linear transformations of PPPs.\\
\\
In this article, we start by presenting a simple model of filtered shot noise process with multiplicative noise and variable input rate (Sec.~\ref{sec:model}). We next consider the general case of PPP transformations and their properties (Sec.~\ref{sec:ppp}). Exact analytic expressions for the joint cumulants of the filtered process are derived (Sec.~\ref{sec:sn-ode-k}) in addition to an approximation of the exact analytical solution (Sec.~\ref{sec:sn-expansion}). Finally, we apply these results to a simple neuronal membrane model of sub-threshold $V_m$ fluctuations with continuously variable rate of presynaptic spikes and explore several practical applications (Sec.~\ref{sec:application}).

\section{Model of Filtered Shot Noise Process}
\label{sec:model}
\begin{figure}[H]
\centering{\includegraphics[width=0.49\columnwidth]{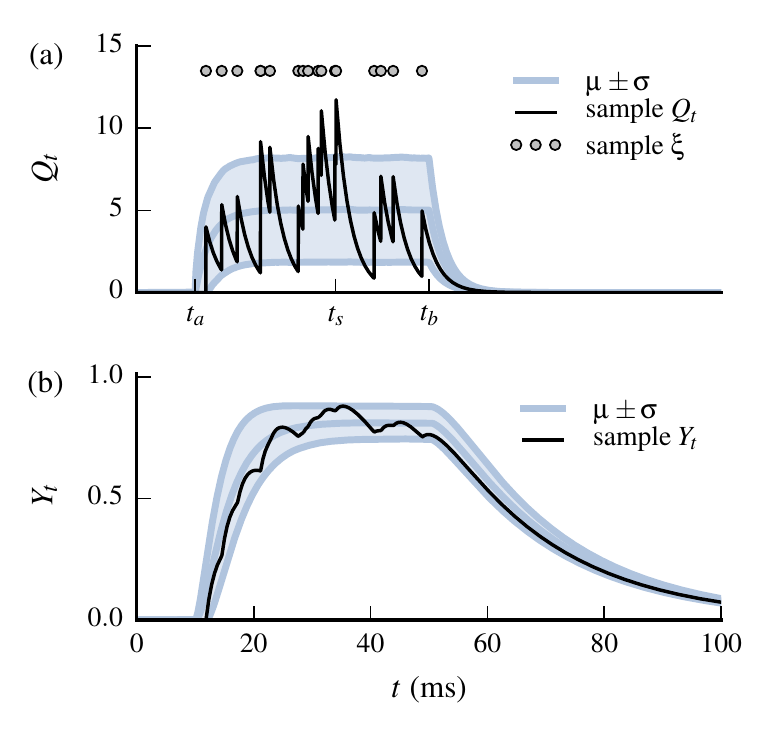}}
\caption{Single realization and basic statistics of filtered shot noise process $Y_t$ under shot noise input $Q_t$. (a) Random input arrival times $x_j\in\xi$ generate nonstationary shot noise $Q_t\equiv Q(t,\xi)$. The input arrival times are distributed with a variable Poisson rate $\lambda(t)$ that restricts the arrivals to occur between $t_a$ and $t_b$. (b) Nonstationary system response $Y_t\equiv Y(t,\xi)$ driven by shot noise $Q_t$. A single realization of random arrival times $\xi$ is represented by gray dots; realizations of $Q_t$ and $Y_t$ are shown in black lines. The mean and standard deviation ($\mu\pm\sigma$) of $Q_t$ and $Y_t$ are shown by gray lines and are clearly nonstationary.}
\label{fig:ode_sn_sim}
\end{figure}

In this section we present a simple model of filtered shot noise process with multiplicative noise. This stochastic process results from the filtering of shot noise input through a linear first-order ODE with variable coefficients. We show that under very simple input rate conditions the filtered process is nonstationary. We derive the time course of the filtered process in terms of the shot noise arrival times. The numerical simulation parameters are presented at the end of this section.\\
\\
Consider the time evolution $Y(t)$ of a system governed by a linear first-order ODE with variable coefficients that is driven by shot noise input $Q(t)$:
\begin{align}
\tau\frac{d}{dt}\,Y(t)&=-Y(t)+\left(1-Y(t)\right)Q(t)\label{eq:ode_sn}\\
\nonumber\\
Q(t)&=\sum_{x_j\in\xi} g(t-x_j)\,H(t-x_j)\label{eq:sn}
\end{align}
where $\tau$ is a time constant, $\xi$ is the set of shot noise arrival times, $g(t-x_j)\,H(t-x_j)$ is the impulse response function at time $t$ for arrival time $x_j\in\xi$ and $H(u)$ is the Heaviside function. The impulse response function is also known as \emph{shot noise kernel}.\\
\\
The input arrival times $\xi$ in Eq.~\eqref{eq:sn} are distributed according to a Poisson law as is characteristic of shot noise. The time evolution of this system is both stochastic and deterministic: stochastic since it is driven by random input arrival times $\xi$, but also deterministic since to each $\xi$ corresponds a unique outcome. The system response $Y(t,\xi)$ is said to be a filtered version of the shot noise process $Q(t, \xi)$ since  Eq.~\eqref{eq:ode_sn} changes its spectral characteristics.\\
\\
Nonstationary dynamics are introduced in the model by restricting input arrival times to occur between $t_a$ and $t_b\geq t_a$ with constant Poisson rate $\lambda$. A single realization of shot noise input $Q(t,\xi)$ and the resulting system response $Y(t,\xi)$ are shown in Fig.~\ref{fig:ode_sn_sim}. The mean and standard deviation ($\mu\pm\sigma$) of both processes are clearly nonstationary since they vary in time.\\
The system response $Y(t,\xi)$ for a particular shot noise input $Q(t,\xi)$ is obtained by solving Eq.~\eqref{eq:ode_sn}. For a given set of input arrival times $\xi$ and initial value $Y_0=0$,
\begin{align}
Y(t,\xi)&=\frac{1}{\tau}\int_{-\infty}^t e^{-\frac{t-z}{\tau}}\,Q(z,\xi)\,e^{-\frac{1}{\tau}\int_z^t Q(u,\xi)\,du}\,dz\nonumber\\
&=\frac{1}{\tau}\int_{-\infty}^t e^{-\frac{t-z}{\tau}}\,\sum_{x_j\in\xi} g(z-x_j)\,H(z-x_j)
\prod_{x_i\in\xi}e^{-\frac{1}{\tau}\int_z^t g(u-x_i)\,H(u-x_i)\,du}\,dz
\label{eq:ode_sol_xi_sm}
\end{align}
\\
The input arrival times $\xi$ completely determine the time evolution of $Y(t,\xi)$. Equation~\eqref{eq:ode_sol_xi_sm} also shows that the response at time $t$ for each input arrival $x_j$ also depends on later input arrivals $x_i\leq t$. For a single shot noise source the solution can be further simplified using integration by parts:
\begin{align}
Y(t,\xi)&=1-\frac{1}{\tau}\int_{-\infty}^t e^{-\frac{t-z}{\tau}}
\prod_{x_i\in\xi}e^{-\frac{1}{\tau}\int_z^t g(u-x_i)\,H(u-x_i)\,du}\,dz\label{eq:ode_sol_xi}
\end{align}
\noindent
The remainder of this article addresses the question of how to obtain the cumulants of the quantity on the left side of Eq.~\eqref{eq:ode_sol_xi} from those on the right side, in the particular case of Poisson distributed input arrival times $\xi$ with variable rate $\lambda(t)$. For reasons of concise presentation, instead of Eq.~\eqref{eq:ode_sol_xi_sm}, we consider the equivalent Eq.~\eqref{eq:ode_sol_xi}.\\
\\
The numerical simulations were generated with the rate function $\lambda(t)$ represented in Fig.~\ref{fig:pp_inh}(a) and exponential kernel shot noise with $g(t-x)=h\exp\left(-(t-x)/\tau_s\right)$. Other parameters are $t_{max}=0.1$ s, $\tau=0.02$ s, $\lambda=500$ Hz, $h=4$ and $\tau_s=0.0025$ s.

\section{Causal Point Process Transformations}
\label{sec:ppp}
\begin{figure}[H]
\centering{
\includegraphics[width=0.49\columnwidth]{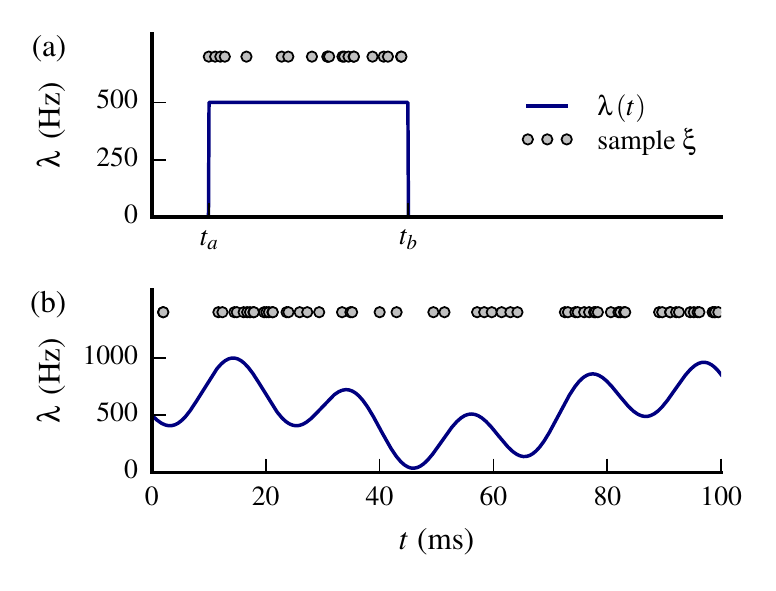}}
\caption{Rate function $\lambda(t)$ examples  for inhomogeneous PPP. A single realization of input arrival times  $\xi$ is represented by gray dots above the rate functions (blue lines) marking the location of input arrival times $x_j\in\xi$. (a) Rate function used to generate input arrival times for the filtered shot noise process of Sec.~\ref{sec:model}. (b) Rate function used to generate presynaptic spike times for the neuronal membrane of Sec.~\ref{sec:application}.}
\label{fig:pp_inh}
\end{figure}

\begin{figure}[H]
\centering{
\includegraphics[width=0.49\columnwidth]{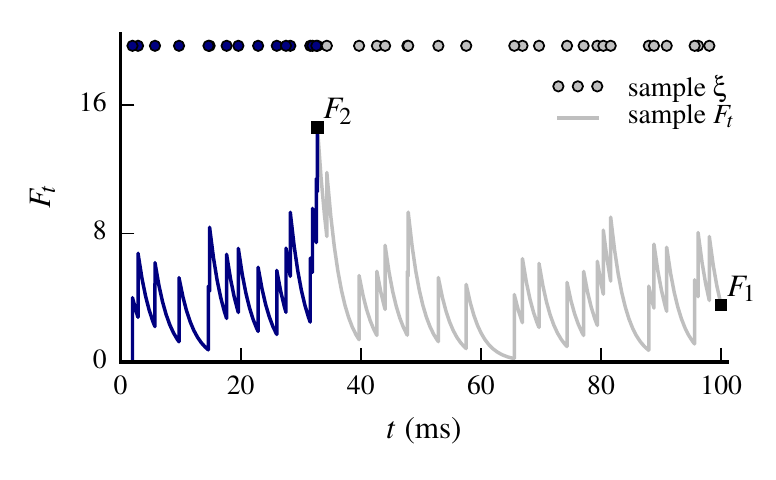}}
\caption{ A shot noise process $F_t$ is a causal PPP transformation $F(t,\xi)$ of input arrival times $x_j\in\xi$. This particular PPP transformation is defined in Eq.~\eqref{eq:sn} and its causality ensures that $F_t$ is not affected by input arrivals later than $t$. For example, the value of $F_2=F(t_2, \xi)$ is determined by input arrivals $x_j\in\xi$ up to $t_2$ (blue line) and is not affected by input arrivals later than $t_2$ (light gray line). The dots above $F_t$ mark the location of input arrival times $x_j\in\xi$.}
\label{fig:sn_sim}
\end{figure}

\begin{figure}[H]
\centering{
\includegraphics[width=0.49\columnwidth]{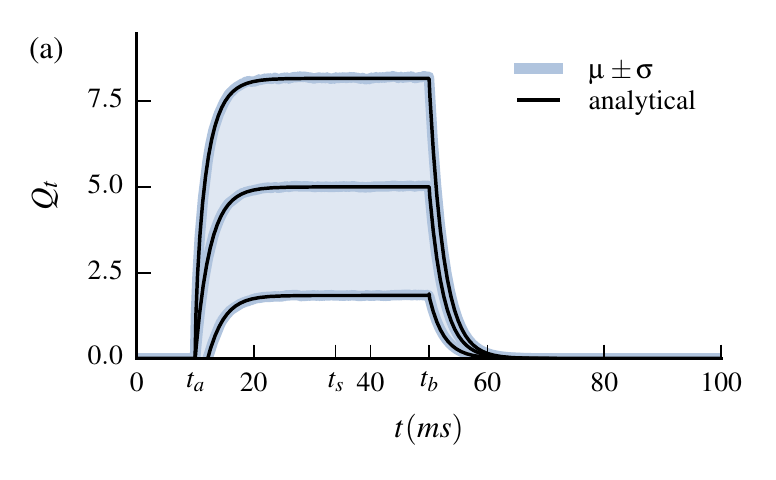}
\includegraphics[width=0.49\columnwidth]{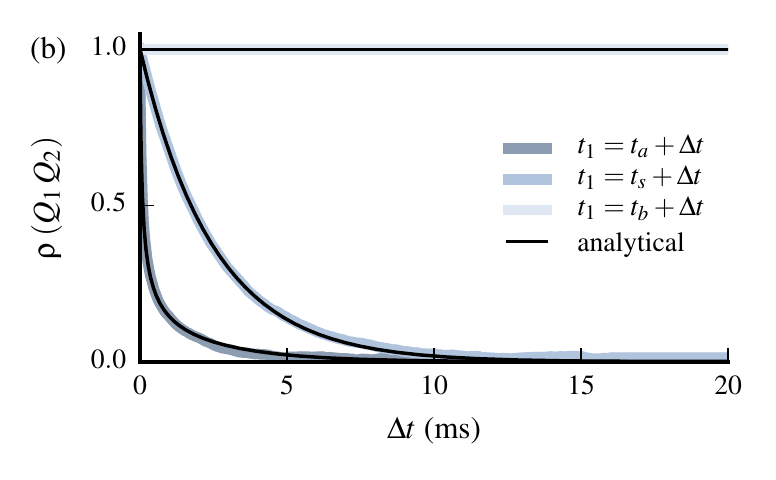}}
\caption{Comparison with numerical simulations for (a) the mean and standard deviation and (b) the autocorrelation of the shot noise process $Q_t$ from Sec.~\ref{sec:model} as predicted by the Campbell Theorem (Eq. \eqref{eq:sn_k}). There is excellent agreement between the simulations (gray lines) and the analytic predictions (black lines) with the respective lines overlapping. The autocorrelation $\rho$ is evaluated at $t_a$, $t_s$, and $t_b$ corresponding respectively to the onset of PPP activity, quasi-stationary $Q_t$, and extinction of PPP activity.}
\label{fig:pred_sn}
\end{figure}

We review the basic properties of PPP transformations and analyze the stochastic process generated by causal PPP transformations. The expectation of PPP transformations yields the joint cumulants of the associated processes. We illustrate this approach with the shot noise process and compare the predicted mean and second order cumulants with numerical simulations.\\
\\
We consider a PPP $\Xi\,(\mathcal{S},\lambda)$ that generates \emph{points} in the interval $\mathcal{S}\subseteq\mathbb{R}$ of the real line with rate function $\lambda(x)\geq 0$ such that $m(\mathcal{S})\equiv\int_{\mathcal{S}}\lambda(x)\,dx$ is finite for any bounded interval $\mathcal{S}$. A realization $\xi$ of $\Xi$ contains a set of $n\geq 0$ points $\{x_1,\ldots,x_n\}\in\mathcal{S}$ that we associate with  input arrival times. A PPP is said to be homogeneous for constant $\lambda(t)=\lambda$ and inhomogeneous otherwise. Example rate functions and sample realizations of the associated inhomogeneous PPP are shown in Fig.~\ref{fig:pp_inh}. These rate functions were used to generate input arrival times for the filtered shot noise process of Sec.~\ref{sec:model} and the presynaptic spikes for the neuronal membrane of Sec.~\ref{sec:application}.\\
\\
We consider a transformation $F(t,\xi)$ that for each real parameter $t\in\mathcal{S}$ and realization $\xi$ evaluates to a positive real number $F_t=F(t,\xi)$. The transformation is assumed invariant under permutation of $x_j\in\xi$, such that when written as a regular function we have $F(t,x_1,\ldots,x_n)=F(t,\{x_1,\ldots,x_n\})$.\\
The expectation of $F(t,\xi)$ is obtained from the ensemble average over the number $n$ of points and their locations  $\{x_1,\ldots,x_n\}$:
\begin{align}
\left\langle F(t,\xi)\right\rangle
&=\sum_{n=0}^\infty \frac{1}{n!}\,e^{-m(\mathcal{S})}\int_\mathcal{S}\cdots\int_\mathcal{S} F\left(t,x_1,\ldots,x_n\right)\prod_{j=1}^n\lambda(x_j)\,dx_j\label{eq:xi_k1}
\end{align}
\\
\noindent
We now focus on the class of PPP transformations that are causal in the time parameter $t$. Such transformations ensure that arrivals $x_j\in\xi$ later than $t$ cannot affect the value of $F(t,\xi)$. A single realization $\xi$ generates the entire time course of $F(t,\xi)$ and we therefore associate a \emph{slave stochastic process} $F_t\equiv F(t,\xi)$ to the causal PPP transformation $F(t,\xi)$. By construction, the expectation of $F_t$ is the expectation of $F(t,\xi)$ given by Eq.~\eqref{eq:xi_k1}. This is illustrated in Fig.~\ref{fig:sn_sim}, where the value of shot noise process $F_t$ at different times is evaluated from the same realization $\xi$.
\noindent
We write $F_1,\ldots,F_K$ for the values of stochastic process $F_t$ at times $t_1,\ldots,t_K$, $\left\langle F_1\cdots F_K\right\rangle$ for its joint moments and $\left\langle\left\langle F_1\cdots F_K\right\rangle\right\rangle$ for its joint cumulants. The expectation of PPP transformations enables to obtain analytical expressions for the joint moments and joint cumulants of $F_t$: its joint moments are obtained by evaluating the expectation of suitable products $F(t_1,\xi)\cdots F(t_K,\xi)$ and its joint cumulants can be constructed explicitly from the joint moments. For example, the moment $\left\langle F_1\,F_2^2\right\rangle$ is evaluated by the expectation $\left\langle F(t_1,\xi)\,F(t_2,\xi)^2\right\rangle$. \\
\\
The causality of $F(t,\xi)$ enables to consider the PPP in the entire real line ($\mathcal{S}=\mathbb{R}$) with finite activity intervals constructed by setting $\lambda(t)=0$ outside the activity windows. This approach yields exact analytical expressions for the joint cumulants of nonstationary processes generated from causal PPP transformations as illustrated next with the nonstationary shot noise process from Sec.~\ref{sec:model}. A shot noise process is a particular type of \emph{random sum}, which is a PPP transformation that factors as $F\left(t,\xi\right)=\sum_{x_j\in\xi} f(t,x_j)$. The joint cumulants of random sums are given by the \emph{Campbell Theorem} \citep{campbell1909study, rice1945mathematical} and are also derived for reference in Appendix~\ref{sec:randomsumproducts}:
\begin{align}
\left\langle\left\langle F_1\ldots F_K\right\rangle\right\rangle
&=\int_\mathcal{S} f(t_1,x)\cdots f(t_K,x)\,\lambda(x)\,dx\label{eq:sn_k}
\end{align}
where $f(t,x)$ is the impulse response function at time $t$ for an input arrival time $x$. Shot noise is a causal random sum with $f(t,x)=g(t-x)\,H(t-x)$.\\
\\
The expectation of more general forms of random sums, such as those in Eq.~\eqref{eq:ode_sol_xi_sm}, are provided by the \emph{Slivnyak-Mecke Theorem} \citep{slivnyak1962some,mecke1962}. A comparison between numerical simulations and \emph{Campbell Theorem} predictions is shown in Fig.~\ref{fig:pred_sn} with excellent agreement for the mean and second order cumulants. The autocorrelation at times $t_1$ and $t_2$ is given by $\rho(F_1\,F_2)=\left\langle\left\langle F_1\,F_2\right\rangle\right\rangle/(\sigma(F_1)\,\sigma(F_2))$ where $\left\langle\left\langle F_1\,F_2\right\rangle\right\rangle$ is the autocovariance at times $t_1$ and $t_2$ and $\sigma(F_t)$ is the standard deviation at time $t$.
\noindent

\section{Exact Analytical Solution}
\label{sec:sn-ode-k}
\begin{figure}[H]
\centering{
\includegraphics[width=0.49\columnwidth]{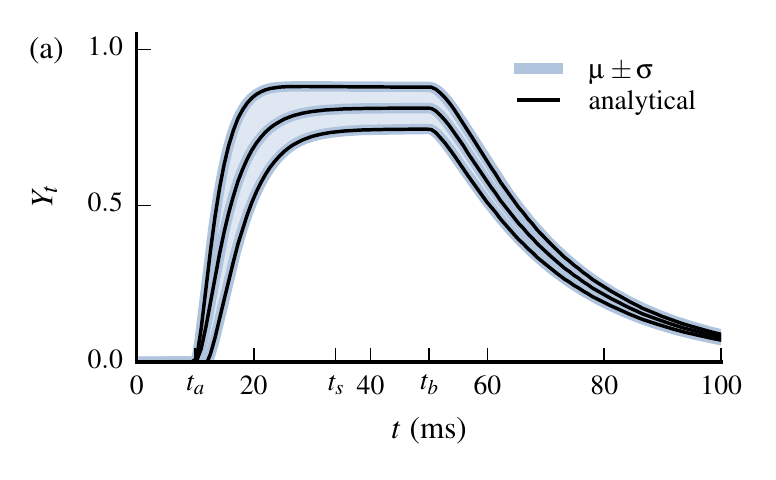}
\includegraphics[width=0.49\columnwidth]{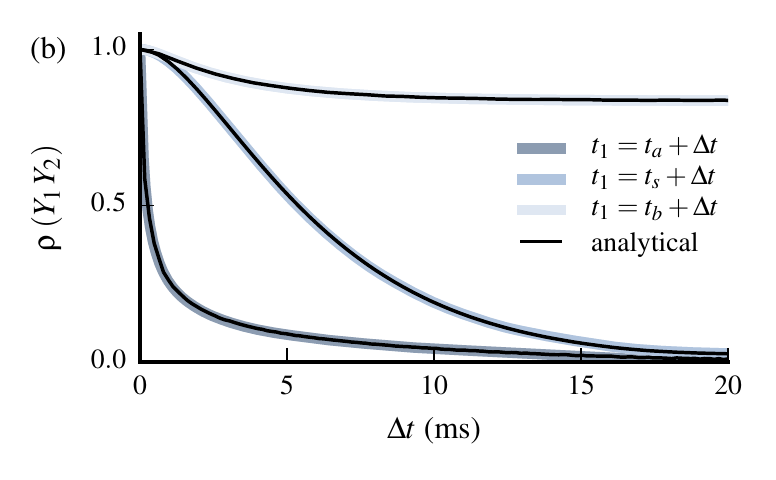}}
\caption{Comparison with numerical simulations for (a) the mean and standard deviation and (b) the autocorrelation of the filtered process $Y_t$ predicted by the exact analytic solution given by Eqs.~\eqref{eq:ode_sn_k1} and \eqref{eq:ode_sn_k}. There is excellent agreement between the simulations (gray lines) and the analytic predictions (black lines) with the respective lines overlapping. The autocorrelation $\rho$ is evaluated at $t_a$, $t_s$, and $t_b$ corresponding respectively to the onset of PPP activity, quasi-stationary $Y_t$, and extinction of PPP activity.}
\label{fig:ode_sn_pred_y}
\end{figure}

We use the properties of PPP transformations to derive exact analytical expressions for the cumulants of filtered shot noise processes with multiplicative noise and variable input rate. We investigate transformations that are relevant to these filtered processes: integral transform and random products. We evaluate their cumulants and compare with numerical simulations the predicted mean and second order cumulants of the filtered process.\\ 
\\
According to Eq.~\eqref{eq:ode_sol_xi}, the filtered process $Y_t$ is the integral of a causal PPP transformation that factors as a product of exponentials of input arrival times $x_j\in\xi$. We now investigate these transformations and define an \emph{integral transform} of $F(t,\xi)$ with regards to a positive and bounded function $w$:
\begin{align}
SF(t,\xi)&=\int_{-\infty}^t F(u,\xi)\,w(u,t)\,du\label{eq:sxi}
\end{align}
\\
The mean and joint moments of the integral transform are calculated by interchanging the infinite sum and integrals of the expectation Eq.~\eqref{eq:xi_k1} with the integral of the transform provided any one side of the equalities is finite (\emph{Fubini-Tonelli Theorem}).
\begin{align}
\left\langle SF_t\right\rangle
&=\left\langle\int_{-\infty}^t F(u,\xi)\,w\left(u,t\right)du\right\rangle
=\int_{-\infty}^t\left\langle F(u,\xi)\right\rangle w\left(u,t\right)du\label{eq:sxi_k1}\\
\left\langle SF_1\cdots SF_K\right\rangle
&=\int_{-\infty}^{t_1}\cdots\int_{-\infty}^{t_K}
\left\langle F(u_1,\xi)\cdots F(u_K,\xi)\right\rangle
\prod_{l=1}^K w\left(u_l,t_l\right)du_l\label{eq:sxi_mean-k}
\end{align}
\\
The linearity of integration extends Eq.~\eqref{eq:sxi_mean-k} to the joint cumulants:
\begin{align}
\left\langle\left\langle SF_1\cdots SF_K\right\rangle\right\rangle
&=\int_{-\infty}^{t_1}\cdots\int_{-\infty}^{t_K}
\left\langle\left\langle F(u_1,\xi)\cdots F(u_K,\xi)\right\rangle\right\rangle 
\prod_{l=1}^K w\left(u_l,t_l\right)du_l\label{eq:sxi_kappa}
\end{align}
\\
We now analyze \emph{random products} that are PPP transformations factoring as $F\left(t,\xi\right)=\prod_{x_j\in\xi} f(t,x_j)$. The joint moments of random products are well known, and as shown in Appendix \ref{sec:randomsumproducts}:
\begin{align}
\left\langle F_1\ldots F_K\right\rangle
=\exp\left(\int_\mathcal{S}\left(\prod_{k=1}^K f(t_k,x)-1\right)\lambda(x)\,dx\right)\label{eq:prod_mean_k}
\end{align}
\\
We have gathered all the elements to derive the mean and joint cumulants of the filtered process $Y_t$. Writing $Q(t,\xi)=Q(t)$ and using the properties of joint cumulants,
\begin{align}
\left\langle Y_t \right\rangle
&=1-\frac{1}{\tau}\int_{-\infty}^t 
 \left\langle e^{-\frac{1}{\tau}\int_{z}^{t} Q(u)\,du}\right\rangle e^{-\frac{t-z}{\tau}}\,dz\label{eq:ode_sn_k1}\\
\left\langle\left\langle Y_1\cdots\,Y_K\right\rangle\right\rangle
&=\left(-\frac{1}{\tau}\right)^K\int_{-\infty}^{t_1}\cdots\int_{-\infty}^{t_K}
\left\langle\left\langle \prod_{k=1}^K e^{-\frac{1}{\tau}\int_{z_k}^{t_k} Q(u)\,du}\right\rangle\right\rangle \prod_{l=1}^K e^{-\frac{t_l-z_l}{\tau}}\,dz_l\label{eq:ode_sn_k}
\end{align}
\\
The expectation of the random product of exponentials is obtained from Eq.~\eqref{eq:prod_mean_k} and yields:
\begin{align}
\left\langle \prod_{k=1}^K e^{-\frac{1}{\tau}\int_{z_k}^{t_k} Q(u)\,du}\right\rangle
&=\exp\left(\int_\mathcal{S}\left(\prod_{k=1}^K e^{-\frac{1}{\tau}\int_{z_k}^{t_k}g(u-x)\,H(u-x)\,du}-1\right)\lambda(x)\,dx\right)\label{eq:expsn_k1}
\end{align}
\\
Replacing Eq.~\eqref{eq:expsn_k1} into Eqs.~\eqref{eq:ode_sn_k1} and \eqref{eq:ode_sn_k} yields the exact solution for the joint cumulants of filtered shot noise process with multiplicative noise and variable input rate. The random product expectation of Eq.~\eqref{eq:expsn_k1} is the key element in the evaluation of the mean and joint cumulants, which was already identified in previous work \citep{wolff2008method,wolff2010mean}, where closed expressions were obtained for expo-
nential kernel shot noise with constant rate. As our derivation shows, this extends to any shot noise kernel $g(t-x)\,H(t-x)$ and variable input rate $\lambda(t)$ and is the main original contribution of this work.\\
\\
The comparison between numerical simulations and the predictions from Eqs.~\eqref{eq:ode_sn_k1} and \eqref{eq:ode_sn_k} are shown in Fig.~\ref{fig:ode_sn_pred_y}. There is excellent agreement even in such a nonstationary scenario with the system undergoing transient evolution. The numerical evaluation of Eqs.~\eqref{eq:ode_sn_k1} and \eqref{eq:ode_sn_k} can be performed very efficiently with the trapezoidal rule due to the double exponential in the integrand.

\section{Central Moments Expansion}
\label{sec:sn-expansion}
\begin{figure}[H]
\centering{
\includegraphics[width=0.49\columnwidth]{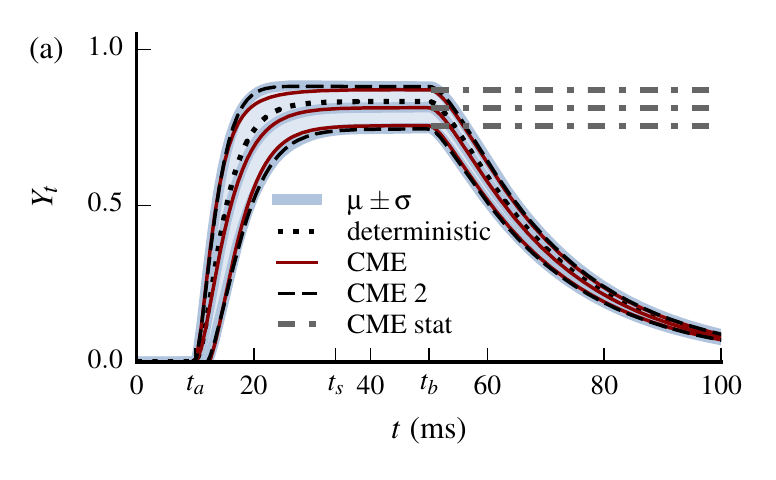}
\includegraphics[width=0.49\columnwidth]{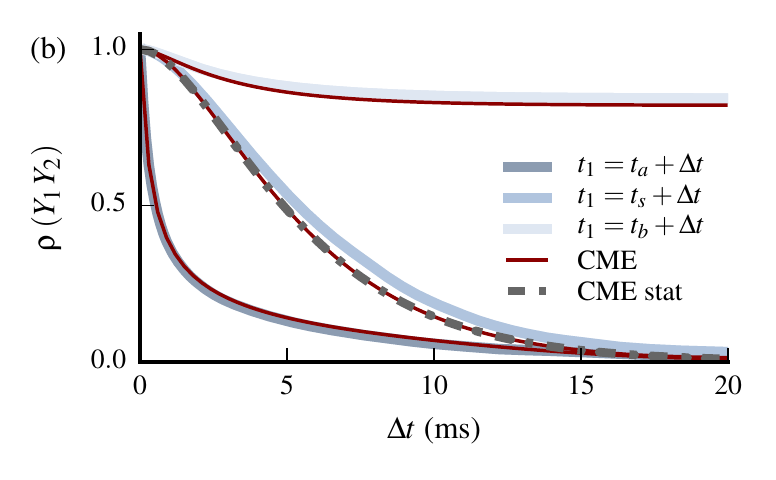}}
\caption{Comparison with numerical simulations for (a) the mean and standard deviation and (b) the autocorrelation of the central moments expansion given by Eqs.~\eqref{eq:ode_sn_approx_k1} and \eqref{eq:ode_sn_approx_k2}. There is good agreement between the simulations (full gray lines) and the approximation (full red lines) for the mean but lower accuracy for second order cumulants. This is corrected by the second order expansion for the autocovariance (dashed black lines) that is provided in Appendix~\ref{sec:cme}. The stationary limit (dashed dotted gray lines) is also shown. The deterministic solution (dotted black line) displays considerable approximation error.}
\label{fig:ode_sn_pred_k1_approx}
\end{figure}

We propose an approximation of the exact analytical solution that is based on a power expansion about the deterministic solution. The central moments expansion (CME) yields a series in the central moments of integrated shot noise. We compare this approximation for the mean and second order cumulants with numerical simulations, including the case of constant Poison rate.\\
\\
The deterministic solution of Eq.~\eqref{eq:ode_sn} with mean shot noise input $\left\langle Q(u)\right\rangle$ is given by:
\begin{align*}
\left\langle Y_t\right\rangle_0
&=1-\frac{1}{\tau}\int_{-\infty}^t e^{-\frac{t-z}{\tau}}
e^{-\frac{1}{\tau}\int_z^t \left\langle Q(u)\right\rangle\,du}\,dz
\end{align*}
\\
This suggests an expansion about the deterministic solution $\left\langle Y_t\right\rangle_0$ by performing power expansions of the random product expectations in Eqs. \eqref{eq:ode_sn_k1} and \eqref{eq:ode_sn_k}. The integrated mean shot noise is first factored out of the random product and a power expansion of the resulting exponential is performed. This corresponds to the delta method technique \citep{cramer1999mathematical,oehlert1992note} for approximating expectations of random variable transformations and yields a series in the central moments of integrated shot noise. As shown in Appendix~\ref{sec:cme}, the second order expansion for a single random product yields:
\begin{align*}
\left\langle e^{-\frac{1}{\tau}\int_z^t Q(u)\,du}\right\rangle
&\simeq e^{-\frac{1}{\tau}\int_z^t \left\langle Q(u)\right\rangle  du}
\left(1+\frac{1}{2}\left\langle\left(-\frac{1}{\tau}\int_z^t\left(Q(u)-\left\langle Q(u)\right\rangle\right)
du\right)^2\right\rangle\right)
\end{align*}
\\
This provides the following approximation for the mean of the filtered process:
\begin{align}
\left\langle Y_t\right\rangle_2
&=1-\frac{1}{\tau}\int_{-\infty}^t
e^{-\frac{1}{\tau}\int_z^t\left(1+\left\langle Q(u)\right\rangle \right) du}
\left(1+\frac{1}{2\tau^2}\int_z^t\int_z^t\left\langle\left\langle Q(u_1)\,Q(u_2)\right\rangle\right\rangle du_1\,du_2\right) dz\label{eq:ode_sn_approx_k1}
\end{align}
\\
where the subscript 2 represents the second order of the expansion.\\
\\
Extending to joint cumulants is straightforward by expanding each exponential individually and collecting terms of same order in $1/\tau$. The first order expansion for the autocovariance is given by:
\begin{align}
\left\langle\left\langle Y_1\,Y_2\right\rangle\right\rangle_1
&=\frac{1}{\tau^4}\int_{-\infty}^{t_1}\int_{-\infty}^{t_2}
e^{-\frac{1}{\tau}\int_{z_1}^{t_1}\left(1+\left\langle Q(u)\right\rangle \right)du-\frac{1}{\tau}\int_{z_2}^{t_2}\left(1+\left\langle Q(v)\right\rangle\right) dv}
\int_{z_1}^{t_1}\int_{z_2}^{t_2}\left\langle\left\langle Q(u_1)\,Q(u_2)\right\rangle\right\rangle du_1\,du_2\,dz_1\,dz_2\label{eq:ode_sn_approx_k2}
\end{align}
\\
The first order expansion for the variance is obtained from Eq.~\eqref{eq:ode_sn_approx_k2} by replacing $t_1=t_2=t$. The comparison between numerical simulations and the predictions from Eqs. \eqref{eq:ode_sn_approx_k1} and \eqref{eq:ode_sn_approx_k2} are shown in Fig.~\ref{fig:ode_sn_pred_k1_approx}. There is good agreement for the mean but lower accuracy for second order cumulants. This can be improved with the second order expansion for the autocovariance that is provided in Appendix~\ref{sec:cme}.
\noindent
We found that the second order expansion for the mean and autocovariance consistently provided good results
in the parameter regimes of neuron cells (as seen in Fig.~\ref{fig:ode_sn_pred_k1_approx}(a) and Fig.~\ref{fig:ode_sn_v_highg_pred} below). Under these conditions, third and fourth order expansions either did not provide significant improvements over the second order or even resulted in worse approximations, in which case much higher order expansions would be required to improve on the second order. Under certain parameter regimes the first order expansion for the autocovariance (Eq.~\eqref{eq:ode_sn_approx_k2}) may already provide good results (see Fig.~\ref{fig:ode_sn_v_pred} below).\\
\\
The stationary limit of the filtered process reflects the statistics of long running trials under shot noise input with constant rate. The cumulants for this regime can be obtained by placing the onset of input arrival times at $-\infty$ and replacing the mean and second order cumulants of shot noise in Eqs~\eqref{eq:ode_sn_approx_k1} and \eqref{eq:ode_sn_approx_k2} with their stationary limits. After integration by parts,
\begin{align}
\left\langle Y_t\right\rangle_2
&=\frac{\left\langle Q\right\rangle}{1+\left\langle Q\right\rangle}
-\frac{\left\langle\left\langle Q^2\right\rangle\right\rangle}{\left(1+\left\langle Q\right\rangle\right)^2}
\frac{1}{\tau}\int_{-\infty}^t e^{-\frac{t-z}{\tau}\left(1+\left\langle Q\right\rangle\right)}\,
r\left(t-z\right) dz\label{eq:ode_sn_approx_stat_k1}\\
\nonumber\\
\left\langle\left\langle Y_1\,Y_2\right\rangle\right\rangle_1
&=\frac{\left\langle\left\langle Q^2\right\rangle\right\rangle}{\left(1+\left\langle Q\right\rangle\right)^2}\frac{1}{\tau^2}
\int_{-\infty}^{t_1}\int_{-\infty}^{t_2}
e^{-\frac{t_1-z_1+t_2-z_2}{\tau}\left(1+\left\langle Q\right\rangle\right)}\,r\left(|z_1-z_2|\right) dz_1\,dz_2\label{eq:ode_sn_approx_stat_k2}
\end{align}
where $\left\langle Q\right\rangle$, $\left\langle\left\langle Q^2\right\rangle\right\rangle$ and $\left\langle\left\langle Q_1\,Q_2\right\rangle\right\rangle=\left\langle\left\langle Q^2\right\rangle\right\rangle r\left(|t_1-t_2|\right)$ are, respectively, the mean, variance, and autocovariance of stationary shot noise.\\
\\
The stationary limits for the mean and second order cumulants of the exponential and alpha kernels are presented in Appendix \ref{sec:staty}.

\section{Application to Neuronal Membranes}
\label{sec:application}
\begin{figure}[H]
\centering{\includegraphics[width=0.49\columnwidth]{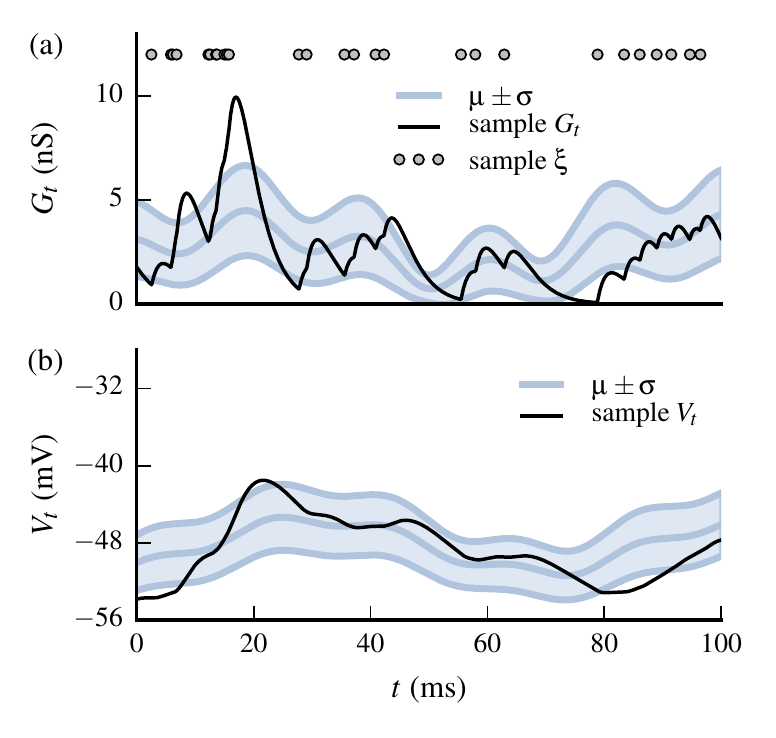}}
\caption{Single realization and basic statistics of membrane potential $V_t$ fluctuations under conductance shot noise input $G_t$. (a) Random presynaptic spike times $x_j\in\xi$ generate nonstationary shot noise conductance $G_t\equiv G(t,\xi)$. The presynaptic spike times are distributed with a continuously varying rate $\lambda(t)$. (b) Nonstationary membrane potential $V_t\equiv V(t,\xi)$ driven by shot noise conductance $G_t$. A single realization of spike times $\xi$ is represented by gray dots; realizations of $G_t$ and $V_t$ are shown in black lines. The mean and standard deviation ($\mu\pm\sigma$) of $G_t$ and $V_t$ are shown with gray lines.}
\label{fig:ode_sn_v_sim}
\end{figure}

We apply the previous results to a simple model of membrane potential $V_m(t)$ fluctuations and explore several practical applications. We first calculate the nonstationary cumulants and compare them with numerical simulations. The central moment expansion (CME) is compared with previously published analytical estimates for the stationary limit of $V_m(t)$. The nonstationary cumulants are integrated in truncated Edgeworth series to approximate the time-evolving distribution of $V_m(t)$, which is compared with numerical simulations. We propose a simple method to estimate $\lambda(t)$ from a small number of noisy realizations of $V_m(t)$ and compare the inferred rate to the original presynaptic rate function. The numerical simulation parameters are presented below.\\
\\
We consider a simple model of the membrane potential $V_m(t)$ for a passive neuron without spiking mechanism that is driven by shot noise conductance $G(t)$. This model has a single synapse type and is directly applicable to experiments where one type of synapse is isolated \citep{okun2008instantaneous}. The time evolution of $V_m(t)$ under conductance shot noise input $G(t)$ is given by the following membrane equation:
\begin{align}
\tau_m\,\frac{d}{dt}\,V_m(t)&=E_l-V_m(t)+\left(E_s-V_m(t)\right)\frac{1}{g_l}\,G(t)\label{eq:ode_sn_vm}\\
\nonumber\\
\frac{1}{g_l}\,G(t)&=\sum_{x_j\in\xi} g(t-x_j)\,H(t-x_j)\label{eq:sn_vm}
\end{align}
where $\tau_m$ is the membrane time constant, $E_l$ is the resting potential, $E_s$ is the synaptic reversal potential, $g_l$ is the leak conductance and $\xi$ is a set of presynaptic spike times.\\
\\
The membrane equation is a scaled and translated version of Eq.~\eqref{eq:ode_sn} with the following change of variables:
\begin{align*}
& V_m(t)=\left(E_s-E_l\right) Y(t)+E_l & Q(t)=\frac{1}{g_l}\,G(t)
\end{align*}
\\
A single realization of conductance shot noise $G_t\equiv G(t,\xi)$ and the resulting membrane potential response $V_t\equiv V_m(t,\xi)$ are shown in Fig.~\ref{fig:ode_sn_v_sim}, where the mean and standard deviation of both processes are also represented. The numerical simulations were generated with the rate function $\lambda(t)$ represented in Fig.~\ref{fig:pp_inh}(b) and alpha kernel shot noise with $g(t-x)=h(t-x/\tau_s)\exp\left(-(t-x)/\tau_s\right)$. Other parameters are $\tau_m=0.02$ s, $E_l=-0.06$ V, $E_s=0$ V and $g_l=10\text{\sc{e}-}9$ S in addition to those detailed in Section \ref{sec:model}. The quantal conductance is $4\text{\sc{e}-}9$ S corresponding to $h=0.4$.

\subsection{Nonstationary Cumulants}
\label{sec:kappa_v}
\begin{figure}[H]
\centering{
\includegraphics[width=0.49\columnwidth]{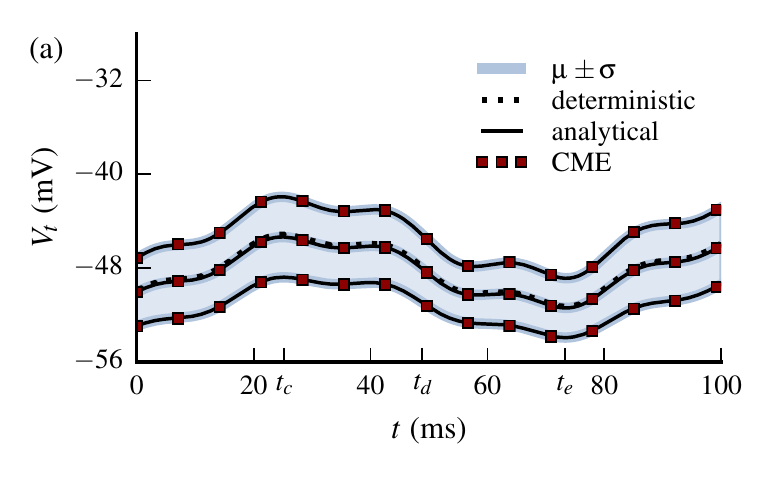}
\includegraphics[width=0.49\columnwidth]{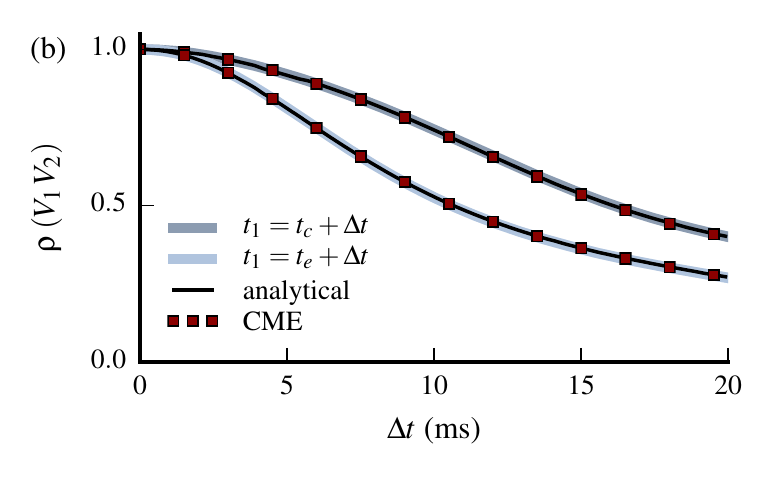}}
\caption{Comparison with numerical simulations for (a) the mean and standard deviation and (b) the autocorrelation of membrane potential $V_t$ predicted by the exact analytical solution (full black lines) and the central moments expansion (red squares). There is excellent agreement between the simulations (full gray lines) and the analytic prediction for both methods with the respective lines overlapping. The deterministic solution (dotted black line) also displays good agreement with mean $V_t$. The autocorrelation $\rho$ is evaluated at local maxima ($t_c$) and minima ($t_e$) of mean $V_t$.}
\label{fig:ode_sn_v_pred}
\end{figure}

\begin{figure}[H]
\centering{
\includegraphics[width=0.49\columnwidth]{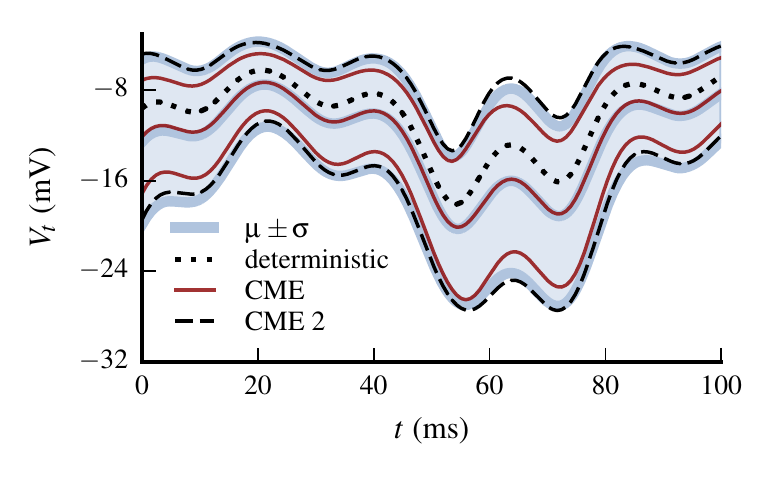}}
\caption{Same parameter regime as Fig.~\ref{fig:ode_sn_v_pred} but with quantal conductance increased by a factor of 20 ($h=80$ nS). The mean and standard deviation of the numerical simulations (full gray lines) display excellent agreement with the exact analytical solution but are omitted for clarity. The approximation error for the CME (full red lines) remains low for the mean but increases significantly for both the standard deviation and the deterministic solution (dotted black line). The second order expansion for the standard deviation (dashed black lines) results in lower approximation error.}
\label{fig:ode_sn_v_highg_pred}
\end{figure}

A first application of this formalism is to derive the mean and joint cumulants of $V_t$ from those of $Y_t$. Using the properties of the mean and cumulants of random variables for each value of $t$ yields the required relationships:
\begin{align}
\left\langle V_t \right\rangle
&=(E_s-E_l)\left\langle Y_t \right\rangle+E_l
&&
\left\langle\left\langle V_1\cdots V_K\right\rangle\right\rangle
=(E_s-E_l)^K\left\langle\left\langle Y_1\cdots Y_K\right\rangle\right\rangle
\label{eq:y2v}
\end{align}
\\
The comparison between numerical simulations and the predictions from Eq.~\eqref{eq:y2v} is shown in Fig.~\ref{fig:ode_sn_v_pred}. There is excellent agreement with the predictions from both the exact analytical solution given by Eqs.~\eqref{eq:ode_sn_k1} and \eqref{eq:ode_sn_k} and the CME given by Eqs~\eqref{eq:ode_sn_approx_k1} and \eqref{eq:ode_sn_approx_k2}. The deterministic solution is obtained from Eq.~\eqref{eq:y2v} by replacing $\left\langle Y_t \right\rangle$ with $\left\langle Y_t \right\rangle_0$ and displays good agreement with mean $V_t$.
\noindent
In this parameter regime, the approximation error of the CME is very low (on the order of $0.01$ mV). However, additional terms of the expansion may be required to reach similar precision in other parameter regimes. In order to illustrate this, we increase  the quantal conductance by a factor of 20 (to 80 nS) with the effect of raising mean $V_t$ very close to the reversal potential $E_s$. As shown in Fig.~\ref{fig:ode_sn_v_highg_pred}, the CME is still in very good agreement for the mean but the approximation error is larger for the standard deviation (on the order of several millivolts). The second order expansion for the standard deviation results in lower approximation error (in the order of 1 mV) but requires evaluating third and fourth order cumulants of integrated shot noise. The approximation error for the deterministic solution also increases to several mV.
\noindent
The Appendix \ref{sec:statv} provides analytical expressions for the CME in the stationary limit of $V_t$ for the mean and second order cumulants of exponential and alpha kernels. These expressions are obtained by applying Eq.~\eqref{eq:y2v} to Eqs. \eqref{eq:ode_sn_approx_stat_k1} and \eqref{eq:ode_sn_approx_stat_k2} and are consistent with previous analytical estimates for the mean and standard deviation that were derived with different approaches: Fokker-Planck methods for exponential kernel shot noise \citep{richardson2005synaptic, rudolph2005extended} given by Eqs. \eqref{eq:ode-sol-vm-stat-exp-k1}, and a shot noise approach for alpha kernel shot noise \citep{kuhn2004neuronal} given by Eqs. \eqref{eq:ode-sol-vm-stat-alpha-k1}. The extension to the autocovariance is given by Eqs. \eqref{eq:ode-sol-vm-stat-exp-k2}, and \eqref{eq:ode-sol-vm-stat-alpha-k2} respectively.

\subsection{Probability Distribution Approximation}
\label{sec:density}
\begin{figure}[H]
\centering{
\includegraphics[width=0.49\columnwidth]{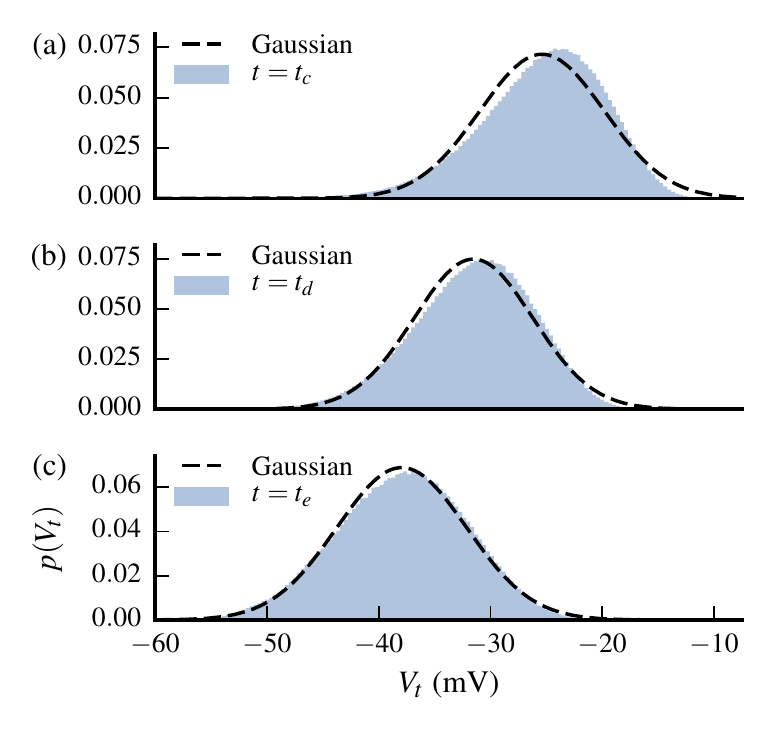}}
\caption{Nonstationary density of membrane potential $p(V_t)$ evaluated at three different times times: $t_c$, $t_d$, and $t_e$ corresponding to (a), (b), and (c), respectively, (see abscissa of Fig.~\ref{fig:ode_sn_v_pred}(a)). Comparison between the empirical histogram (gray) and the Gaussian approximation (dashed line) based on nonstationary mean and variance of $V_t$. The time evolution of $p(V_t)$ is captured successfully by this approximation, which as expected also misses the skew of $p(V_t)$.}
\label{fig:density_v_gauss}
\end{figure}

\begin{figure}[H]
\centering{
\includegraphics[width=0.49\columnwidth]{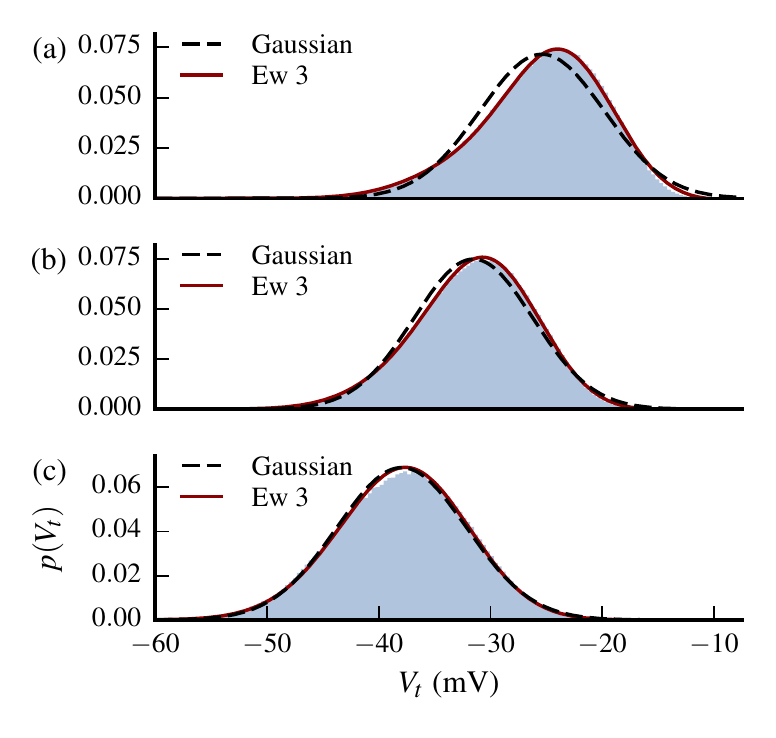}
\includegraphics[width=0.49\columnwidth]{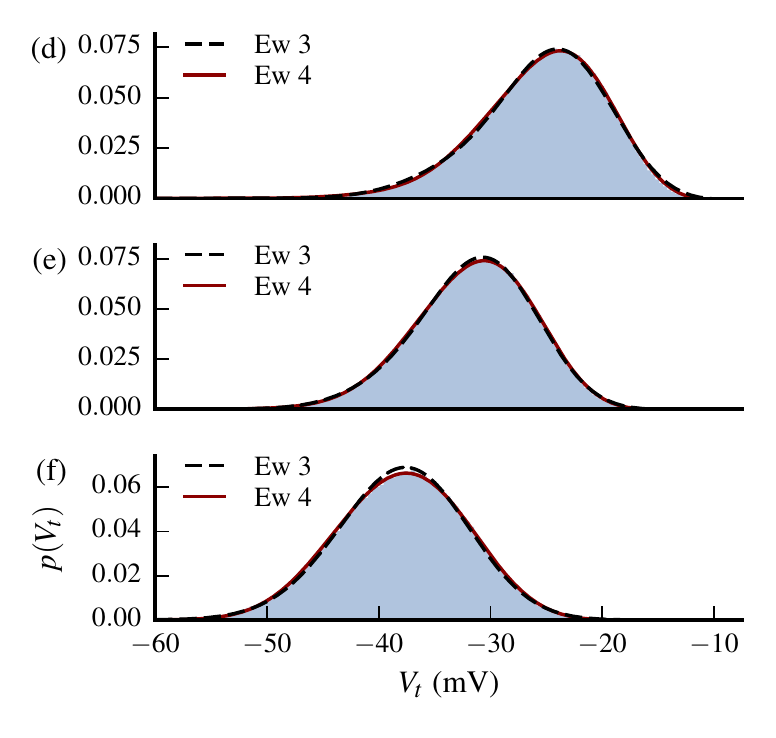}}
\caption{(a)-(c): Comparison at times $t_c$, $t_d$, and $t_e$ between the empirical histogram (gray), Gaussian approximation (dashed black line), and the third order Edgeworth series (full red line). (d)-(f): Comparison at times $t_c$, $t_d$, and $t_e$ between the empirical histogram (gray), third order Edgeworth series (dashed black line), and fourth order Edgeworth series (full red line). The slight discrepancy at the peak of the empirical histogram in (c) is successfully captured by the fourth order Edgeworth series in (f).}
\label{fig:density_v_hw3}
\end{figure}

A second application of this formalism is to use the nonstationary cumulants to approximate the time evolving distribution of membrane potential fluctuations. The mean and standard deviation of $V_t$ yield a Gaussian approximation that successfully captures the time evolution of $p(V_t)$ as illustrated in Fig.~\ref{fig:density_v_gauss}. As expected, the skew of the distribution is not well captured by the Gaussian approximation, which has been reported in both experimental \citep{destexhe1999impact} and theoretical studies \citep{richardson2005synaptic, rudolph2005extended}. The quantal conductance was increased by a factor of 4 (to 16 nS) in these simulations.\\
\\
Deviations from the Gaussian distribution are expected whenever cumulants of order three or higher are present in $p(V_t)$. We use a truncated Edgeworth series \citep{edgeworth1907representation,cramer1999mathematical,wallace1958asymptotic} to account for these deviations since it provides an asymptotic expansion of $p(V_t)$ in terms of its cumulants. In particular, we use the Edgeworth series expanded from the Gaussian distribution distribution as discussed in \citep{badel2011firing}. This has the advantage of coinciding with the Gaussian approximation whenever cumulants of order three or higher are negligible. This is an important aspect since approximately Gaussian shapes of $p(V_t)$ are sometimes present in experimental intracellular recordings. In terms of the normalized process $X_t=\left(V_t-\left\langle V_t\right\rangle\right)/\sigma_t$ with $\sigma_t\equiv\sqrt{\left\langle\left\langle V_t^2\right\rangle\right\rangle}$, the truncated fourth order Edgeworth series is given by:
\begin{align}
p_{Ew 4}\left(X_t=x\right)
&=\frac{1}{\sigma_t}\left(1+\frac{1}{3!}\frac{\left\langle\left\langle V_t^3\right\rangle\right\rangle}{\sigma_t^3}\left(x^3-3x\right)\right.\nonumber\\
&\qquad\qquad
\left.+\frac{1}{4!}\frac{\left\langle\left\langle V_t^4\right\rangle\right\rangle}{\sigma_t^4}\left(x^4-6x^2+3\right)
+\frac{10}{6!}\frac{\left\langle\left\langle V_t^3\right\rangle\right\rangle^2}{\sigma_t^6}\left(x^6-15x^4+45x^2-15\right)\right)
\mathcal{N}(x)
\end{align}
where $\mathcal{N}(x)=\exp(-x^2/2)/\sqrt{2\pi}$ is the standard normal density and $p(V_t=v)\simeq p_{Ew}\left(x=\frac{v-\left\langle V_t\right\rangle}{\sigma_t}\right)$. The third order Edgeworth series is given by the the first two terms.\\
\noindent
As illustrated in the left side of Fig.~\ref{fig:density_v_hw3}, the skewness of $p(V_t)$ is indeed captured by the third order of the Edgeworth series. Figure \ref{fig:density_v_hw3}(c) also shows a slight overestimation near the peak of $p(V_t)$, which is successfully captured by the fourth order, as shown in the right side of this figure.
\noindent
Under more extreme parameter regimes, additional terms of the Edgeworth series may be needed to approximate $p(V_t)$. In such cases, the asymptotic character of the series becomes relevant since the truncation error is of the same order as the first neglected term of the series. An important caveat is that the truncated series may yield negative values for certain values of $x$. This is intrinsic to Edgeworth series that are constructed in the set of orthogonal polynomials associated with the base distribution (\emph{Hermite polynomials} in the case of the standard normal distribution). The truncated series integrates to unity but may result in an invalid density function since negatives values are possible. Algorithms for computing an Edgeworth series to an arbitrary order are provided in Refs.~\citep{petrov1962vestnik,blinnikov1998expansions}.

\subsection{presynaptic Rate Estimation}
\label{sec:extract}
\begin{figure}[H]
\centering{
\includegraphics[width=0.49\columnwidth]{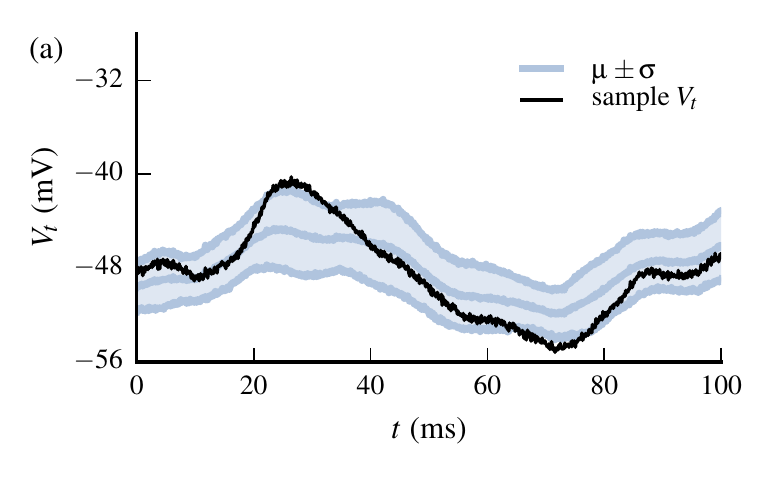}
\includegraphics[width=0.49\columnwidth]{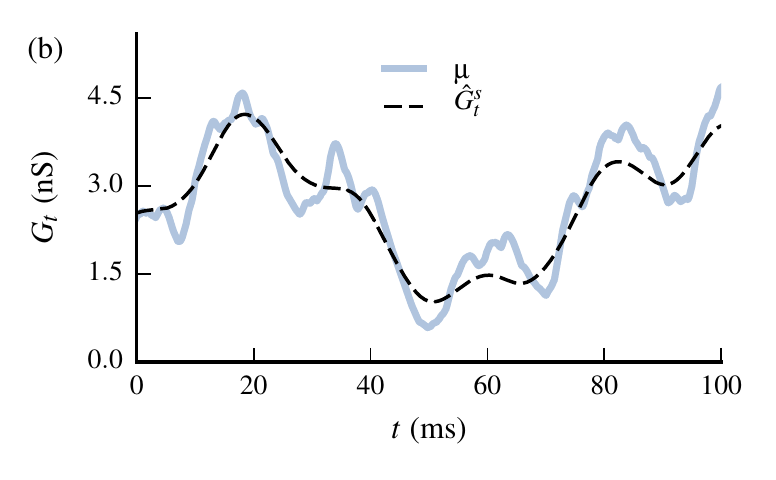}}
\caption{(a) Single realization (black line) and basic statistics (gray lines) generated from a small number ($N=10$) of membrane potential $V_t$ traces with additive noise to simulate measurement error. (b) Extracted mean conductance (full gray line) and smoothed version $G_t^s$ (dashed black line) obtained with non-parametric smoothing.}
\label{fig:extract1}
\end{figure}

\begin{figure}[H]
\centering{
\includegraphics[width=0.49\columnwidth]{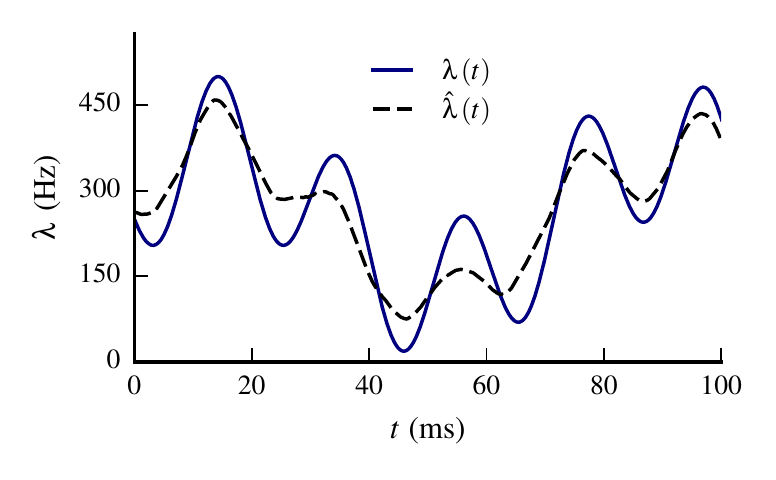}}
\caption{Estimated presynaptic rate $\hat{\lambda}(t)$ (black dashed line) compared with original rate function $\lambda(t)$ (full blue line). The magnitude and variations are reasonably well captured considering the small number ($N = 10$) of membrane potential traces used.}
\label{fig:extract_2}
\end{figure}

Another application of this formalism is to estimate the nonstationary presynaptic rate $\lambda(t)$ from a small number ($N=10$) of membrane potential $V_t$ traces that are independently generated from the same PPP. Each trace has small amounts of additive noise to simulate measurement error that are independent from the PPP. The traces of $V_t$ are sampled at rate $1/\Delta t$. A single realization of the noisy membrane potential with mean and variance estimated from a small number of traces is shown in Fig.~\ref{fig:extract1}. The noisy membrane equation is given by:
\begin{align}
\tau_m\,\frac{d}{dt}\,V_m(t)&=E_l-V_m(t)+\left(E_s-V_m(t)\right)\frac{1}{g_l}\,G(t) + \epsilon(t)\label{eq:ode_sn_vm_noisy}
\end{align}
where $\epsilon(t)$ is a zero mean Gaussian white noise in units of voltage with $\sigma(\epsilon)=0.2$ mV.\\
\\
The key expression that enables to estimate $\lambda(t)$ from traces of $V_t$ is the Campbell Theorem for the mean of nonstationary shot noise given by Eq.~\eqref{eq:sn_k}. If the shot noise kernel is known then the rate function can in principle be obtained by deconvolution of the mean conductance. However, this operation is very sensitive to noise since small changes in the estimated mean conductance will result in large changes of the estimated rate function. This aspect is dealt with by smoothing the estimated mean conductance prior to performing the deconvolution step. From each trace of $V_t$ we extract the input conductance by inverting Eq.~\eqref{eq:ode_sn_vm_noisy} and average them to obtain the estimated mean conductance:
\begin{align*}
\left\langle \hat{G}_t\right\rangle
&=\frac{1}{N}\sum_{n=1}^N \frac{g_l}{\Delta t}\,\frac{\tau_m\left(V_{t+\Delta t}^n-V_t^n\right)-\Delta t\left(E_l-V_t^n\right)}{E_s-V_t^n}
\end{align*}
where $V_t^n$ is the $n$-th trace of $V_t$ and $\Delta t$ is the sampling interval.\\
\noindent
The estimated mean conductance will be a noisy version of the actual mean conductance due to the effects of the additive noise $\epsilon(t)$ in each trace of $V_t$. Non-parametric smoothing is performed using a local linear smoother with tricube kernel and kernel bandwidth selected by cross-validation \citep{wasserman2006all}, yielding the smoothed version $\left\langle \hat{G}_t^s\right\rangle$ shown in Fig.~\ref{fig:extract1}(b). Finally, we use the discrete convolution theorem to estimate the presynaptic rate $\hat{\lambda}(t)$ from the smoothed mean conductance:
\begin{align*}
\hat{\lambda}(t)&=\frac{1}{\Delta t}\,\text{DFT}^{-1}\left[\frac{\text{DFT}\left\{\left\langle \hat{G}_t^s\right\rangle\right\}}{\text{DFT}\{g\}}\right]
\end{align*}
\\
The result is shown in Fig.~\ref{fig:extract_2}, where the estimated $\hat{\lambda}(t)$ rate compares favorably to the original presynaptic rate $\lambda(t)$. Estimating $\lambda(t)$ from noisy shot noise data has been previously addressed \citep{sequeira1995intensity} in addition to methods that enable to estimate the shot noise kernel \citep{sequeira1997blind}.

\section{Discussion}
\label{sec:discussion}
In this paper, we investigated important statistical properties
of filtered shot noise processes with multiplicative noise, in the
general case of variable input rate. These properties provide a
compact description of time-evolving dynamics of the system.
Such processes arise from the filtering of nonstationary shot
noise input through a linear first-order ODE with variable
coefficients. We have obtained general results for this class
of stochastic processes and results specific to applications in
neuronal models.\\
\\
We first identified the causal PPP transformation that
corresponds to filtered shot noise with multiplicative noise. We
investigated the statistical properties of this transformation to
derive the exact analytical solution for the joint cumulants
of the filtered process with variable input rate. Excellent
agreement with numerical simulations was found for the mean
and second order cumulants. We proposed an approximation
based on a CME about the solution of the deterministic
system. We have shown with numerical simulations that under
parameter regimes relevant to neuronal membranes the second
order of this approximation provides good results for the mean
and second order cumulants. Under certain parameter regimes,
the first order expansion for the second order cumulants may
already provide effective approximations.\\
\\
These general results were then applied to a simple model
of subthreshold membrane potential $V_m$ fluctuations subject
to shot noise conductance with continuously variable rate
of presynaptic spikes. Excellent agreement with numerical
simulations was found for the mean and second order
cumulants for both the exact analytical solution and the
second order CME. This approximation is consistent with
previously published analytical estimates for stationary $V_m$ .
An expression for the stationary limit of autocovariance is
provided for exponential and alpha kernel shot noise input.
An approximation for the time-evolving distribution of $V_m$
is proposed that is based on a truncated Edegeworth series
using the nonstationary cumulants obtained analytically. This
approximation successfully captures the time evolution of $V_m$
under a large range of conditions. The nonstationary mean
of shot noise is used to estimate the presynaptic rate from
a small number of intracellular $V_m$ recordings with additive
noise simulating measurement error.\\
\\
In future work we will extend this formalism to multiple
independent shot noise inputs by applying the Slivnyak-Mecke
Theorem. Such development would yield direct applications
for neuronal membrane models with different synapse types
(such as excitatory and inhibitory synapses). Preliminary work
indicates that analytic treatment of filtered shot noise with
correlated input is accessible with this formalism. The present
work also opens perspectives for the analytical development
first passage time statistics based on nonstationary approxima-
tions of the filtered process distribution.

\subsection*{Acknowledgments}
Research was supported by the CNRS, the Agence Nationale de la Recherche (ANR; ComplexV1 project), and the European Community (BrainScales FP7-269921 and the
Human Brain Project FP7-604102). M.B. was supported by a PhD fellowship from the European Marie-Curie Program (FACETS-ITN FP7-237955).

\appendix

\section*{Appendix}

\section{Random Sums and Random Products}
\label{sec:randomsumproducts}
Random products are transformations of PPP that factor as $F\left(t,\xi\right)=\prod_{x_j\in\xi} f(t,x_j)$. The expectation of random products is obtained as follows:
\begin{align*}
\left\langle F_t\right\rangle
&=\sum_{n=0}^\infty \frac{1}{n!}\,e^{-m(\mathcal{S})}\left(\int_\mathcal{S} f(t,x)\,\lambda(x)\,dx\right)^n=\exp\left(\int_\mathcal{S}\left(f(t,x)-1\right)\lambda(x)\,dx\right)\nonumber\\
\left\langle F_1\ldots F_K\right\rangle
&=\left\langle \prod\limits_{x_j\in\xi}\prod\limits_{k=1}^K f(t_k,x_j)\right\rangle
=\exp\left(\int_\mathcal{S}\left(f(t_1,x)\cdots f(t_K,x)-1\right)\lambda(x)\,dx\right)
\end{align*}
\\
In the case of the random product with $f(t,x_j)=e^{-\frac{1}{\tau}\int_z^t g(u-x_j)H(u-x_j)du}$ and  $\mathcal{S}=\mathbb{R}$,
\begin{align*}
\left\langle e^{-\frac{1}{\tau}\int_z^t Q(u,\xi)\,du}\right\rangle
&=\exp\left(\int_{-\infty}^z \left(e^{-\frac{1}{\tau}\int_z^t g(u-x)\,du}-1\right)\lambda(x)\,dx
+\int_z^t \left(e^{-\frac{1}{\tau}\int_y^t g(v-y)\,dv}-1\right)\lambda(y)\,dy\right)
\end{align*}
\\
Random sums are transformations of PPP that factor as $F\left(t,\xi\right)=\sum_{x_j\in\xi} f(t,x_j)$. The joint cumulants of random sums are given by the \emph{Campbell Theorem} \citep{campbell1909study, rice1945mathematical}. The \emph{characteristic function} $\phi(s_1, \ldots, s_K)$ is the expectation of a random product, and its derivatives yield the joint cumulants:
\begin{align*}
\phi(s_1, \ldots, s_K)&\equiv\left\langle e^{is_1F(t_1,\xi)+\cdots+is_KF(t_K,\xi)}\right\rangle
=\left\langle \prod_{x_j\in\xi}\prod_{k=1}^K e^{is_k f(t_k,x_j)}\right\rangle
=\exp\left(\int_\mathcal{S}\left(e^{\sum_{k=1}^Kis_k f(t_k,x)}-1\right)\lambda(x)\,dx\right)
\end{align*}
\begin{align*}
\left\langle\left\langle F(t_1,\xi)\ldots F(t_K,\xi)\right\rangle\right\rangle
&=\left.\left(\frac{1}{i}\,\frac{d}{ds_1}\right)\cdots\left(\frac{1}{i}\,\frac{d}{ds_K}\right)
\ln\phi(s_1, \ldots, s_K)\right|_{s_1,\ldots,s_K=0}
=\int_\mathcal{S} f(t_1,x)\cdots f(t_K,x)\,\lambda(x)\,dx
\end{align*}

\section{Central Moments Expansion}
\label{sec:cme}
A Taylor expansion of the random product about mean shot noise input results in a series of central moments of the integrated shot noise. Expanding the exponential inside the expectation, keeping terms of order $(1/\tau_m)^2$ and re-expressing in terms of cumulants, yields:
\begin{align*}
\left\langle e^{-\frac{1}{\tau}\int_z^t Q(u,\xi)\,du}\right\rangle
&=e^{-\frac{1}{\tau}\int_z^t \left\langle Q(u)\right\rangle du}\left(1+\sum_{m=2}^{+\infty}\frac{1}{m!}\left\langle \left(-\frac{1}{\tau}\int_z^t \left(Q(u,\xi)-\left\langle Q(u)\right\rangle\right)\,du\right)^m\right\rangle\right)\\
&\simeq e^{-\frac{1}{\tau} S\bar{Q}}\left(1
+\frac{1}{2\,\tau^2}\left\langle\left\langle SQ^2\right\rangle\right\rangle\right)
\end{align*}
where $SQ^2\equiv\int_z^t Q^2(v,\xi)\,dv$ and $S\bar{Q}\equiv\int_z^t\left\langle Q(u)\right\rangle\,du$.\\
\\
Higher order cumulants are obtained in a similar manner by expanding each exponential individually and collecting terms in the same order of $1/\tau_m$. The second order expansion for second order cumulants involves the expansion of two random products and keeping terms up to order $(1/\tau_m)^4$, yielding:
\begin{align*}
&\left\langle\left\langle e^{-\frac{1}{\tau}\int_{z_1}^{t_1} Q(u_1,\xi)\,du_1}\,e^{-\frac{1}{\tau}\int_{z_2}^{t_2} Q(u_2,\xi)\,du_2}\right\rangle\right\rangle\\
&\quad 
\simeq e^{-\frac{1}{\tau}S\bar{Q}_1-\frac{1}{\tau}S\bar{Q}_2}
\Bigg(\frac{1}{\tau^2}\left\langle\left\langle SQ_1\,SQ_2\right\rangle\right\rangle
-\frac{1}{2\tau^3}\left(
\left\langle\left\langle SQ_1^2\,SQ_2\right\rangle\right\rangle
+\left\langle\left\langle SQ_1\,SQ_2^2\right\rangle\right\rangle\right)\\
&\qquad\qquad\qquad\qquad\quad
+\frac{1}{2\tau^4}\Big(
\frac{1}{3}\left\langle\left\langle SQ_1^3\,SQ_2\right\rangle\right\rangle
+\frac{1}{3}\left\langle\left\langle SQ_1\,SQ_2^3\right\rangle\right\rangle
+\frac{1}{2}\left\langle\left\langle SQ_1^2\,SQ_2^2\right\rangle\right\rangle\\
&\qquad\qquad\qquad\qquad\qquad\qquad
+\left\langle\left\langle SQ_1\,SQ_2\right\rangle\right\rangle
\left(\left\langle\left\langle SQ_1^2\right\rangle\right\rangle
+\left\langle\left\langle SQ_2^2\right\rangle\right\rangle
+\left\langle\left\langle SQ_1\,SQ_2\right\rangle\right\rangle\right)\Big)\Bigg)
\end{align*}
where $SQ_1\equiv\int_{z_1}^{t_1} Q(v,\xi)\,dv$, $SQ_1^2\equiv\int_{z_1}^{t_1} Q^2(v,\xi)\,dv$ and $S\bar{Q}_1\equiv\int_{z_1}^{t_1}\left\langle Q(u)\right\rangle\,du$, etc.\\

\subsection{Stationary limit for $Y_t$}
\label{sec:staty}
The stationary limit of shot noise autocovariance can be written $\left\langle\left\langle Q_1\,Q_2\right\rangle\right\rangle=\left\langle\left\langle Q^2\right\rangle\right\rangle r\left(|t_1-t_2|\right)$, since:
\begin{align*}
\left\langle\left\langle Q_1\,Q_2\right\rangle\right\rangle
&=\lambda\int_{-\infty}^{\min(t_1,t_2)} g(t_1-x)\,g(t_2-x)\,dx
=\lambda\int_0^{+\infty} g(u)\,g(|t_1-t_2|+u)\,du
=\left\langle\left\langle Q^2\right\rangle\right\rangle r\left(|t_1-t_2|\right)
\end{align*}
with $r\left(|t_1-t_2|\right)\equiv\int_0^{+\infty} g(u)\,g(|t_1-t_2|+u)\,du/\int_0^{+\infty} g(v)^2\,dv$.\\
\\
For the exponential kernel shot noise $r\left(|t_1-t_2|\right)=e^{-\frac{|t_1-t_2|}{\tau_s}}$ and the stationary mean and second order cumulants are given by:
\begin{align*}
\left\langle Q\right\rangle
&=\lambda h\tau_s
&\left\langle\left\langle Q_1\,Q_2\right\rangle\right\rangle
&=\frac{\lambda h^2\tau_s}{2}\,e^{-\frac{|t_1-t_2|}{\tau_s}}
=\left\langle\left\langle Q^2\right\rangle\right\rangle\,e^{-\frac{|t_1-t_2|}{\tau_s}}
\end{align*}
\\
Writing $Q_0\equiv 1+\left\langle Q\right\rangle$ and applying Eq.~\eqref{eq:ode_sn_approx_stat_k1} yields the mean:
\begin{align*}
\left\langle Y_t\right\rangle_2
&=\frac{\left\langle Q\right\rangle}{Q_0}
-\frac{\left\langle\left\langle Q^2\right\rangle\right\rangle}{Q_0^2}\frac{1}{\tau}
\int_{-\infty}^t e^{-\frac{t-z}{\tau}\,Q_0}\,e^{-\frac{t-z}{\tau_s}}\,dz
=\left\langle Y_t \right\rangle_0
-\frac{\left\langle\left\langle Q^2\right\rangle\right\rangle}{Q_0^2\left(Q_0+\frac{\tau}{\tau_s}\right)}
\end{align*}
\\
Applying Eq.~\eqref{eq:ode_sn_approx_stat_k2} yields the autocovariance:
\begin{align*}
\left\langle\left\langle Y_1\,Y_2\right\rangle\right\rangle_1
&=\frac{\left\langle\left\langle Q^2\right\rangle\right\rangle}{Q_0^2}
\frac{1}{\tau^2}\int_{-\infty}^{t_1}\int_{-\infty}^{t_2}
e^{-\frac{t_1-z_1+t_2-z_2}{\tau}\,Q_0}\,e^{-\frac{|z_1-z_2|}{\tau_s}} dz_1\,dz_2\\
&=\left\{ 
  \begin{array}{l l}
  \frac{\left\langle\left\langle Q^2\right\rangle\right\rangle}{Q_0^2\left(Q_0+\frac{\tau}{\tau_s}\right)\left(Q_0-\frac{\tau}{\tau_s}\right)}
\left(e^{-\frac{|t_1-t_2|}{\tau_s}}
-\frac{1}{Q_0}\frac{\tau}{\tau_s}
\,e^{-\frac{|t_1-t_2|}{\tau}\,Q_0}\right) & \quad \text{if }Q_0\neq \frac{\tau}{\tau_s}\\
\\
\frac{\left\langle\left\langle Q^2\right\rangle\right\rangle}{2\,\tau\,Q_0^3}\left(\tau_s+|t_1-t_2|\right) e^{-\frac{|t_1-t_2|}{\tau_s}} & \quad \text{otherwise}
\end{array} \right.
\end{align*}
\\
Setting $t_1=t_2=t$ in the previous result yields the variance:
\begin{align*}
\left\langle\left\langle Y_t^2\right\rangle\right\rangle_1
&=\frac{\left\langle\left\langle Q^2\right\rangle\right\rangle}{Q_0^3\left(Q_0+\frac{\tau}{\tau_s}\right)}
\end{align*}
\\
For the alpha kernel shot noise $r(|t_1-t_2|)=e^{-\frac{|t_1-t_2|}{\tau_s}}\left(1+\frac{|t_1-t_2|}{\tau_s}\right)$ and the stationary mean and second order cumulants are given by:
\begin{align*}
\left\langle Q\right\rangle
&=\lambda h\tau_s
&\left\langle\left\langle Q_1\,Q_2\right\rangle\right\rangle
&=\frac{\lambda h^2\tau_s}{4}\,e^{-\frac{|t_1-t_2|}{\tau_s}}\left(1+\frac{|t_1-t_2|}{\tau_s}\right)
=\left\langle\left\langle Q^2\right\rangle\right\rangle\,e^{-\frac{|t_1-t_2|}{\tau_s}}
\left(1+\frac{|t_1-t_2|}{\tau_s}\right)
\end{align*}
\\
Proceeding as before yields:
\begin{align*}
\left\langle Y_t\right\rangle_2
&=\frac{\left\langle Q\right\rangle}{Q_0}
-\frac{\left(Q_0+2\,\frac{\tau}{\tau_s}\right)\left\langle\left\langle Q^2\right\rangle\right\rangle}{Q_0^2\left(Q_0+\frac{\tau}{\tau_s}\right)^2}
&&
\left\langle\left\langle Y_t^2\right\rangle\right\rangle_1
=\frac{\left(Q_0+2\,\frac{\tau}{\tau_s}\right)\left\langle\left\langle Q^2\right\rangle\right\rangle}{Q_0^3\left(Q_0+\frac{\tau}{\tau_s}\right)^2}
\end{align*}
\begin{align*}
\left\langle\left\langle Y_1\,Y_2\right\rangle\right\rangle_1
&=\left\{ 
  \begin{array}{l l}
  \frac{\left\langle\left\langle Q^2\right\rangle\right\rangle}{Q_0^2\left(Q_0+\frac{\tau}{\tau_s}\right)^2\left(Q_0-\frac{\tau}{\tau_s}\right)^2}\\
\times\Bigg(\left( \left(1+\frac{|t_1-t_2|}{\tau_s}\right) \left(Q_0+\frac{\tau}{\tau_s}\right)\left(Q_0-\frac{\tau}{\tau_s}\right)-2\left(\frac{\tau}{\tau_s}\right)^2\right)e^{-\frac{|t_1-t_2|}{\tau_s}}\\
\qquad\qquad\qquad\qquad\qquad\qquad\qquad\qquad\qquad\qquad
+\frac{2}{Q_0}\left(\frac{\tau}{\tau_s}\right)^3
\,e^{-\frac{|t_1-t_2|}{\tau}Q_0}\Bigg) & \text{if }Q_0\neq \frac{\tau}{\tau_s}\\
\\
\frac{\left\langle\left\langle Q^2\right\rangle\right\rangle}{4\,\tau\,Q_0^3}\left(3(\tau_s+|t_1-t_2|)+\frac{1}{\tau_s}\,|t_1-t_2|^2\right) e^{-\frac{|t_1-t_2|}{\tau_s}} & \text{otherwise}
\end{array} \right.
\end{align*}

\subsection{Stationary limit for $V_t$}
\label{sec:statv}
We apply the transformation given by Eq.~\eqref{eq:y2v} to the results from the previous section to obtain the cumulants for the membrane potential $V_t$. The expression for the stationary mean of the deterministic system is the same for both shot noise kernels:
\begin{align*}
\left\langle V_t\right\rangle_0
&=\frac{\left\langle G\right\rangle}{G_0}\left(E_s-E_l\right) + E_l
=\frac{g_l\,E_l+\left\langle G\right\rangle E_s}{G_0}
&& G_0=g_l+\left\langle G\right\rangle
\end{align*}
\\
The mean and variance of exponential and alpha kernel shot noise are consistent with those given in Refs.~\citep{richardson2005synaptic, rudolph2005extended}  and Ref.~\citep{kuhn2004neuronal}, respectively. The extension to the autocovariance is also provided below. For exponential kernel shot noise and using $E_e-E_l = \frac{G_0}{g_l}\left(E_e-\left\langle V_t \right\rangle_0\right)$,
\begin{align}
\left\langle V_t\right\rangle_2
&=\left\langle V_t\right\rangle_0
-\frac{\left\langle\left\langle G^2\right\rangle\right\rangle}{g_l\left(\frac{G_0}{g_l}+\frac{\tau}{\tau_s}\right)G_0}\left(E_s-\left\langle V_t\right\rangle_0\right)
&&\left\langle\left\langle V_t^2\right\rangle\right\rangle_1
=\frac{\left\langle\left\langle G^2\right\rangle\right\rangle}{g_l\left(\frac{G_0}{g_l}+\frac{\tau}{\tau_s}\right) G_0}
\left(E_s-\left\langle V_t\right\rangle_0\right)^2\label{eq:ode-sol-vm-stat-exp-k1}
\end{align}
\begin{align}
\left\langle\left\langle V_1\,V_2\right\rangle\right\rangle_1
&=\left\{ 
  \begin{array}{l l}
  \frac{\left\langle\left\langle G^2\right\rangle\right\rangle}{g_l^2\left(\frac{G_0}{g_l}+\frac{\tau}{\tau_s}\right)\left(\frac{G_0}{g_l}-\frac{\tau}{\tau_s}\right)}
\left(e^{-\frac{t_1-t_2}{\tau_s}}
-\frac{\tau}{\tau_s}\frac{g_l}{G_0}
\,e^{-\frac{t_1-t_2}{\tau}\frac{G_0}{g_l}}\right)
\left(E_s-\left\langle V_t\right\rangle_0\right)^2 & \quad \text{if }\frac{G_0}{g_l}\neq\frac{\tau}{\tau_s}\\
\\
\frac{\left\langle\left\langle G^2\right\rangle\right\rangle}{2\,\tau\,g_l\,G_0}\left(\tau_s+|t_1-t_2|\right) e^{-\frac{|t_1-t_2|}{\tau_s}}\left(E_s-\left\langle V_t\right\rangle_0\right)^2 & \quad \text{otherwise}
\end{array} \right.\label{eq:ode-sol-vm-stat-exp-k2}
\end{align}
\\
For alpha kernel shot noise,
\begin{align}
\left\langle V_t\right\rangle_2
&=\left\langle V_t\right\rangle_0
-\frac{\left(\frac{G_0}{g_l}+2\,\frac{\tau}{\tau_s}\right)\left\langle\left\langle G^2\right\rangle\right\rangle}{g_l\left(\frac{G_0}{g_l}+\frac{\tau}{\tau_s}\right)^2 G_0}
\left(E_s-\left\langle V_t\right\rangle_0\right)
&&\left\langle\left\langle V_t^2\right\rangle\right\rangle_1
=\frac{\left(\frac{G_0}{g_l}+2\,\frac{\tau}{\tau_s}\right)\left\langle\left\langle G^2\right\rangle\right\rangle}{g_l\left(\frac{G_0}{g_l}+\frac{\tau}{\tau_s}\right)^2 G_0}
\left(E_s-\left\langle V_t\right\rangle_0\right)^2\label{eq:ode-sol-vm-stat-alpha-k1}
\end{align}
\begin{align}
\left\langle\left\langle V_1\,V_2\right\rangle\right\rangle_1
&=\left\{ 
  \begin{array}{l l}
  \frac{\left\langle\left\langle G^2\right\rangle\right\rangle}{g_l^2\left(\frac{G_0}{g_l}+\frac{\tau}{\tau_s}\right)^2\left(\frac{G_0}{g_l}-\frac{\tau}{\tau_s}\right)^2}
\left(E_s-\left\langle V_t\right\rangle_0\right)^2\\
\times
\Bigg(\left(\left(1+\frac{t_1-t_2}{\tau_s}\right)
\left(\frac{G_0}{g_l}+\frac{\tau}{\tau_s}\right)
\left(\frac{G_0}{g_l}-\frac{\tau}{\tau_s}\right)-2\left(\frac{\tau}{\tau_s}\right)^2\right)e^{-\frac{t_1-t_2}{\tau_s}}\\
\qquad\qquad\qquad\qquad\qquad\qquad\qquad\qquad\qquad\qquad
+2\left(\frac{\tau}{\tau_s}\right)^3\frac{g_l}{G_0}
\,e^{-\frac{t_1-t_2}{\tau}\frac{G_0}{g_l}}\Bigg) & \text{if }\frac{G_0}{g_l}\neq\frac{\tau}{\tau_s}\\
\\
\frac{\left\langle\left\langle G^2\right\rangle\right\rangle}{4\,\tau\,g_l\,G_0}\left(3(\tau_s+|t_1-t_2|)+\frac{1}{\tau_s}\,|t_1-t_2|^2\right)  e^{-\frac{|t_1-t_2|}{\tau_s}}\left(E_s-\left\langle V_t\right\rangle_0\right)^2 & \text{otherwise}
\end{array} \right.\label{eq:ode-sol-vm-stat-alpha-k2}
\end{align}


\bibliography{bibliography}
\bibliographystyle{plainnat}

\end{document}